\documentclass[12pt]{article}
\usepackage{graphicx}
\usepackage{epsfig}
\begin {document}
\begin{center}
{\bf PARTICLE PRODUCTION IN THE CENTRAL REGION}

{\bf AT LHC ENERGIES}

\vspace{.2cm}

C. Merino$^*$, C. Pajares$^*$, and Yu.M. Shabelski$^{**}$ \\

\vspace{.5cm}
$^*$ Departamento de F\'\i sica de Part\'\i culas, Facultade de F\'\i sica, \\ 
and Instituto Galego de F\'\i sica de Altas Enerx\'\i as (IGFAE), \\ 
Universidade de Santiago de Compostela, \\
E-mail: merino@fpaxp1.usc.es \\
E-mail: pajares@fpaxp1.usc.es \\

\vspace{.2cm}

$^{**}$ Petersburg Nuclear Physics Institute, \\
E-mail: shabelsk@thd.pnpi.spb.ru \\
\vskip 0.9 truecm

{\it Contribution to the Proceedings of Low-$x$ Meeting 2011\\
Santiago de Compostela, Galice, Spain, 3$^{rd}$-7$^{th}$ June 2011}\\
\vspace{0.4cm}

{\it Talk presented by Yu.M. Shabelski}

\vspace{1.2cm}

A b s t r a c t
\end{center}

We consider the first LHC data for $pp$ collisions in the framework of 
 Regge phenomenology, and the Quark-Gluon String Model, and we present 
the corresponding predictions for both the integral cross sections and 
the inclusive densities of secondaries, that are determined by Pomeron 
exchange. All parameters were fixed long time ago for the description of 
the data at fixed target energies. The first measurements of the total 
inelastic cross section by ATLAS and CMS Collaborations are in agreement 
with our calculations. Also the inclusive densities measured in the 
central region are in agreement with our theoretical predictions on 
the accuracy level of about 10$-$15\%.

\vskip 1.5cm


\newpage

\section{Introduction}

In Regge theory the Pomeron exchange dominates the high energy soft hadron
interaction, the contributions of all other exchanges to the total or
inelastic cross sections becoming negligibly small at LHC energies.

The Quark-Gluon String Model (QGSM) \cite{KTM} is based on Dual Topological 
Unitarization (DTU), Regge phenomenology, and nonperturbative notions of QCD.
This model is successfully used for the description of multiparticle production 
processes in hadron-hadron \cite{KaPi,Sh,Ans,ACKS,AMPS,MRS,MPRS}, 
hadron-nucleus \cite{KTMS,Sh1}, and nucleus-nucleus \cite{JDDS} collisions. 
In particular, the inclusive densities of different secondaries produced in 
$pp$ collisions at $\sqrt{s} = 200$~GeV in midrapidity region were 
reasonably described in ref. \cite{AMPS} (see also refs.~\cite{MRS,MPRS}).

In the QGSM high energy interactions are considered as proceeding via the 
exchange of one or several Pomerons, and all elastic and inelastic processes
result from cutting through or between Pomerons \cite{AGK}. Inclusive spectra 
of hadrons are related to the corresponding fragmentation functions of quarks 
and diquarks, which are constructed using the Reggeon counting rules \cite{Kai}.
The quantitative predictions of the QGSM depend on the values of several
parameters that were fixed at once by the comparison of the corresponding
calculations to the experimental data obtained at fixed target energies.

The experimental data obtained at LHC allow us to test the stability of 
the QGSM predictions and of the values of the model parameters. Fortunately,
one can see that the model predictions are in reasonable agreement with the
first LHC data.

In this paper we discuss the cross sections, inclusive densities of
secondaries, and the antibaryon/baryon ratios at LHC energies in the 
framework of QGSM. 

\section{Cross sections}

For the Pomeron trajectory
\begin{equation}
\alpha_P(t) = 1 + \Delta + \alpha'_P \cdot t\;,\,\, \Delta > 0 \;,
\end{equation}
the one-Pomeron contribution to $\sigma^{tot}_{hN}$ equals
\begin{equation}
\sigma_P = 8\pi\cdot\gamma\cdot e^{\Delta \cdot \xi}, \;
{\rm with}\; \xi = \ln s/s_0 \;,
\end{equation}
where $\gamma = g_1(0) \cdot g_2(0)$ is the Pomeron coupling, 
$s_0 \simeq 1$ GeV$^2$, and $\sigma_P$ rises with energy as $s^{\Delta}$. 
The correct asymptotic behavior $\sigma^{tot}_{hN} \sim \ln^2s$, compatible 
with the Froissart bound~\cite{Froi}, can be obtained by taking into account 
the multipomeron exchanges in the diagrams of Fig.~1.
\begin{figure}[htb]
\centering
\vskip -4.cm
\mbox{\psfig{file=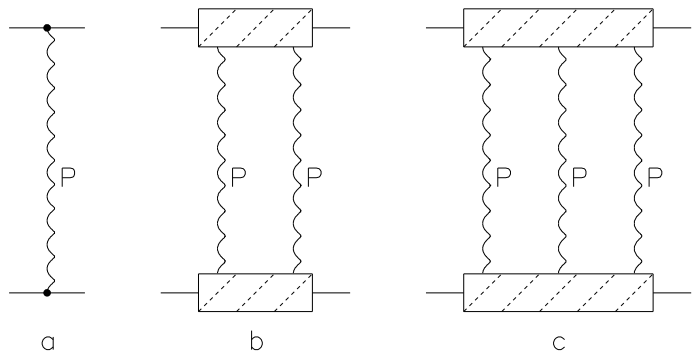,width=0.65\textwidth}} \\
\vskip -.3cm
\caption{\footnotesize
Regge-pole theory diagrams: (a) single, (b) double, (c) and triple
Pomeron exchange in elastic $hN$ scattering. The hadron couplings
to two and three Pomerons are shown by dashed boxes.}
\end{figure}

A simple quasi-eikonal treatment \cite{Kar3} yields to
\begin{equation}
\sigma^{tot}_{hN} = \sigma_P\cdot f(z/2) \;,\,\, \; \sigma^{el}_{hN} =
\frac{\sigma_P}{C}\cdot [f(z/2) - f(z)] \;,
\end{equation}
\begin{equation}
f(z)  =  \sum^{\infty}_{k=1} \frac1{k \cdot k !}\cdot (-z)^{k-1} =
\frac1z \int ^z_0 \frac{dx}x\cdot (1-e^{-x}) \;, 
\end{equation}
\begin{equation}
z = \frac{2C\cdot\gamma}{\lambda}\cdot e^{\Delta \xi} \;,\,\,
\lambda = R^2 + \alpha'_P\cdot\xi \;.
\end{equation}
Here, $R^2$ is the radius of the Pomeron and $C$ is the quasi-eikonal
enhancement coefficient.

At asymptotically high energies ($z \gg 1$) we obtain
\begin{equation}
\sigma^{tot}_{hN} = \frac{8\pi\cdot \alpha'_P\cdot\Delta}{C}\cdot\xi^2 \;,\,\,
\;
\sigma^{el}_{hN} = \frac{4\pi\cdot\alpha'_P\cdot\Delta}{C^2}\cdot\xi^2 \;,
\end{equation}
according to the Froissart limit \cite{Froi}.

The numerical values of the Pomeron parameters were taken \cite{Sh} to be :
\begin{equation}
\Delta = 0.139 , \;\;  \alpha'_P = 0.21 \, {\rm GeV}^{-2}, \;\;
\gamma = 1.77 \, {\rm GeV}^{-2}, \;\; R^2 = 3.18 \, {\rm GeV}^{-2}, 
\;\; C = 1.5. 
\end{equation}

The predictions of Regge theory obtained wity these values of the parameters 
(and accounting for a small contribution from non-Pomeron exchange) are
presented in Table.~1.

\begin{center}

\vskip 5pt

\begin{tabular}{|c||r|r|r|r|r|} \hline
$\sqrt{s}$  & $\sigma^{tot}$ & $\sigma^{el}$ & $\sigma^{inel}$ & 
$\sigma^{inel}$(ATLAS~\cite{ATLAS}) & $\sigma^{inel}$(CMS~\cite{CMS1}) 
\\ \hline
900 GeV & 67.4 & 13.2 & 54.2 & & \\
7 TeV   & 94.5 & 21.1 & 73.4 & $69.4\pm 2.4$\hspace{0.75cm} & 
66.8 - 74.8\hspace{0.5cm}  \\
14 TeV & 105.7 & 24.2 & 81.5 & & \\ \hline
\end{tabular}

\vskip 5pt

\end{center}
Table 1. {\footnotesize The Regge theory predictions for total, total
elastic, and total inelastic cross sections (in mb) in $pp$ collisions
at LHC energies.} 

\vskip 5.pt

These results are in agreement with those in ref.~\cite{AAPS}. 

The experimental points for $\sigma^{inel}_{pp}$ measured by ATLAS and CMS
Collaboration~\cite{ATLAS,CMS1} presented in Table 1 (we omit the error bar
coming from extrapolation) are in a reasonable agreement with our
calculations.

However, in the complete Reggeon diagram technique~\cite{Grib} not only
Regge-poles and cuts but also more complicated diagrams (e.g. the so-called
enhanced diagrams) should be taken into account. In the numerical calculations
of such diagrams some new uncertainties appear due to the fact that the
vertices of the coupling of $n$ and $m$ Reggeons are unknown. 

The numerical calculations which account for enhanced diagrams 
\cite{KMR,GL,Ost} lead the values of $\sigma^{inel}_{pp}$ of the same 
order ($\pm$ 10\% at $\sqrt{s} = 14$ TeV) as those presented in Table~1.
The values of $\sigma^{inel}_{pp}$ calculated in refs. \cite{KMR,GL} are 
slightly smaller than the experimental values \cite{ATLAS,CMS1}, and while
two sets of calculations in ref.~\cite{Ost} lead to too large cross
section, the third one~\cite{Ost} practically coincides with our results.

\section{Inclusive densities}

The Quark-Gluon String Model (QGSM) \cite{KTM,KaPi,Sh} allows us to make 
quantitative predictions of different features of multiparticle production, 
in particular, the  inclusive densities of different secondaries both in the 
central and in the beam fragmentation regions. In QGSM high energy
hadron-nucleon collisions are considered as taking place via the exchange of
one or several Pomerons, all elastic and inelastic processes resulting from
cutting through or between Pomerons~\cite{AGK}. 

Each Pomeron corresponds to a cylindrical diagram (see Fig.~2a), and thus, when 
cutting one Pomeron, two showers of secondaries are produced as it is shown in 
Fig.~2b. The inclusive spectrum of a secondary hadron $h$ is then determined 
by the convolution of the diquark, valence quark, and sea quark distributions 
$u(x,n)$ in the incident particles, with the fragmentation functions $G^h(z)$ 
of quarks and diquarks into the secondary hadron $h$. These distributions, as 
well as the fragmentation functions, are constructed by using the Reggeon
counting rules~\cite{Kai}. Both the diquark and the quark distribution
functions depend on the number $n$ of cut Pomerons in the considered diagram.

\begin{figure}[htb]
\centering
\includegraphics[width=.6\hsize]{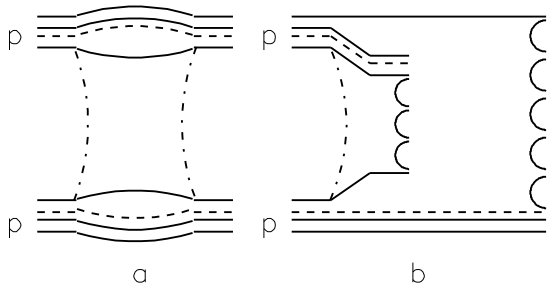}
\caption{\footnotesize
(a) Cylindrical diagram corresponding to the one--Pomeron exchange 
contribution to elastic $pp$ scattering, and (b) the cut of this diagram which 
determines the contribution to the inelastic $pp$ cross section. Quarks 
are shown by solid curves and string junction by dashed curves.}
\end{figure}

For a nucleon target, the inclusive rapidity $y$ (or Feynman-$x$ $x_F$) 
spectrum of a secondary hadron $h$ has the form~\cite{KTM}:
\begin{equation}
\frac{dn}{dy}\ = \
\frac{x_E}{\sigma_{inel}}\cdot \frac{d\sigma}{dx_F}\ = \frac{dn}{dy} =
\sum_{n=1}^\infty w_n\cdot\phi_n^h (x) +  w_D\cdot\phi_D^h (x) \ ,
\end{equation}
where the functions $\phi_{n}^{h}(x)$ determine the contribution of the 
diagram with $n$ cut Pomerons and $w_n$ is the relative weight of this 
diagram. The last term $w_D\cdot\phi_D^h (x)$ accounts for the 
contribution of diffraction dissociation processes.

For $pp$ collisions
\begin{eqnarray}
\phi_{pp}^h(x) & = & f_{qq}^{h}(x_+,n)\cdot f_q^h(x_-,n) +
f_q^h(x_+,n)\cdot f_{qq}^h(x_-,n) + \nonumber\\
& + &  2\cdot(n-1)\cdot f_s^h(x_+,n)\cdot f_s^h(x_-,n)\ ,
\end{eqnarray}

\begin{equation}
x_{\pm} = \frac12\cdot\left[\sqrt{4m_T^2/s+x^2}\ \pm x\right] ,
\end{equation}
where $f_{qq}$, $f_q$, and $f_s$ correspond to the contributions of diquarks, 
valence quarks, and sea quarks, respectively.

These functions are determined by the convolution of the diquark and quark 
distributions (that are normalized to one) with the fragmentation functions,
e.g. for the quark one can write:
\begin{equation}
f_q^h(x_+,n)\ =\ \int\limits_{x_+}^1u_q(x_1,n)\cdot G_q^h(x_+/x_1) dx_1\ .
\end{equation}

At very high energies both $x_+$ and $x_-$ are negligibly small in the 
midrapidity region, and so all fragmentation functions, which are usually 
written \cite{Kai} as $G^h_q(z) = a_h\cdot(1-z)^{\beta}$, become constants and 
equal for a particle and its antiparticle: 
\begin{equation}
G_q^h(x_+/x_1) = a_h \ . 
\end{equation}
This leads, in agreement with~\cite{AKM}, to
\begin{equation}
\frac{dn}{dy}\ = \ g_h \cdot (s/s_0)^{\alpha_P(0) - 1}
\sim a^2_h \cdot (s/s_0)^{\alpha_P(0) - 1} \,,
\end{equation}
that corresponds to the only one-Pomeron exchange diagram (AGK theorem 
\cite{AGK}) at asymptotically high energy. The values of the Pomeron 
parameters presented in Eq.~(7) are used in the QGSM numerical 
calculations. 

The values $dn/dy$ in the central region can be obtained under
different conditions. Sometimes they are presented for all inelascic 
interactions, while in other cases they correspond to the events
without single diffraction (NSD), or to events called INEL$> 0$, in which 
as minimum one charged particle should be detected in the kinematical 
window $\vert \eta \vert > 1$~\cite{ALICE}. When considering the experimental
value $dn/dy = N_{particles}/N_{events}$ one has to keep in mind that though
for both NSD and INEL$> 0$ triggers the value of $N_{particles}$ in
midrapidity region at LHC energies is constant, the number of events
$N_{events}$ significantly changes in these three cases, leading to different
values of $dn/dy$.

As an example, we present in Table~2 three values of $dn/dy$ measured by the 
ALICE Collaboration~\cite{ALIC,ALICE,ALICE1} at $\sqrt{s} = 900$ GeV. 

\begin{center}

\vskip 5pt

\begin{tabular}{|c||r|r|r|} \hline
& All inelastic \cite{ALIC} & NSD~\cite{ALIC}\hspace{0.5cm}
& INEL$> 0$ \cite{ALICE,ALICE1}  
\\ \hline
dn/dy & $3.02 \pm 0.01^{0.08}_{0.05}$\hspace{0.3cm} &  $3.58 \pm 0.01^{0.12}_{0.12}$ & 
 $3.81 \pm 0.01^{0.07}_{0.07}$\hspace{0.4cm}  \\ \hline
\end{tabular}
\end{center}
Table 2. {\footnotesize The inclusive densities of charged secondaries
measured by ALICE Collaboration in all inelastic events,
in events without single diffraction (NSD), and in events with as minimum
one charged particle in the kinematical window $\vert \eta \vert > 1$
(INEL$> 0$), in $pp$ collisions at $\sqrt{s} = 900$
GeV~\cite{ALIC,ALICE,ALICE1}.}

\vskip 5pt

The QGSM predictions for the $dn/d\eta$ distributions of all charged 
secondaries produced in inelastic $pp$ and $\bar{p}p$ collisions at 
different energies are shown in Fig.~3. The experimental data are taken 
from ref. \cite{Kaid}.

\begin{figure}[htb]
\centering
\vskip -1.cm
\includegraphics[width=.65\hsize]{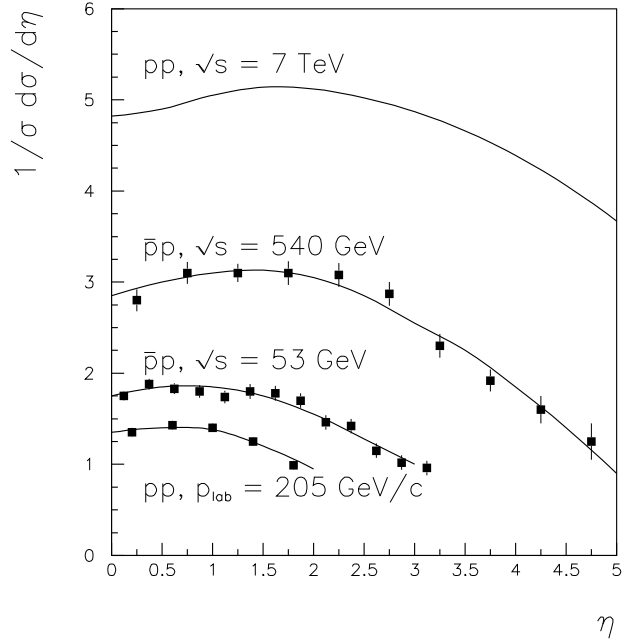}
\vskip -.7cm
\caption{\footnotesize 
The QGSM predictions for the pseudorapidity distributions of all charged 
secondaries produced in inelastic $pp$ and $\bar{p}p$ collisions at 
different energies.}
\end{figure}

The QGSM allows one to calculate the inclusive spectra of different 
secondaries. The comparison of our predictions~\cite{MPRS} 
with new data by the ALICE Collaboration \cite{ALICE2} is presented in Table 3. 

\begin{center}

\vskip 5pt

\begin{tabular}{|c||r|r|} \hline
Particle & QGSM \cite{MPRS} & ALICE Collaboration \cite{ALICE2} \\   \hline
$\pi^+$         & 1.68\hspace{0.6cm} & $1.493 \pm 0.004 \pm 0.074$\hspace{0.3cm}  \\ 
$\pi^-$         & 1.66\hspace{0.6cm} & $1.485 \pm 0.004 \pm 0.074$\hspace{0.3cm}  \\
$K^+$           & 0.17\hspace{0.6cm} & $0.184 \pm 0.004 \pm 0.015$\hspace{0.3cm}  \\
$K^-$           & 0.16\hspace{0.6cm} & $0.183 \pm 0.004 \pm 0.015$\hspace{0.3cm}  \\
$\overline{p}$  & 0.10\hspace{0.6cm} & $0.077 \pm 0.002 \pm 0.006$\hspace{0.3cm}   \\
$\overline{\Lambda}$  & 0.05\hspace{0.6cm} & 0.08\hspace{3.4cm}     \\
$\overline{\Xi^+}$    & 0.005\hspace{0.4cm} & 0.009\hspace{3.2cm}   \\
$\overline{\Omega^+}$ & 0.0004\hspace{0.2cm} & 0.0008\hspace{3.cm}   \\
\hline
\end{tabular}

\end{center}
Table 3. {\footnotesize The QGSM predictions \cite{MPRS} for
the midrapidity yields $dn/dy$ ($\vert y \vert < 0.5$) of
different secondaries at energy $\sqrt{s} = 900$ GeV together
with the corresponding experimental data by the ALICE
Collaboration \cite{ALICE2}.}

\vskip 5pt

The inclusive densities of pions are slightly overestimated. The 
predicted~\cite{MPRS} value of the ratio of $\Xi^-/\Lambda$ midrapidity yields
($\sim 0.10$) is in agreement with the experimental result~\cite{ALICE3}
 $\Xi^-/\Lambda \simeq 0.11 \pm 0.005$. The agreement of the order of 
10$-$15\% for all cases can be considered as a reasonable one.

We predict \cite{MPRS} for all inelastic interactions one increase of the 
midrapidity yields for $\pi$ and $K$ mesons of about 1.4 times in the 
energy region $\sqrt{s} = 900$ GeV$-$7 TeV (smaller than the increase in 
the data \cite{CMS} for $K^0_s$ equal to $1.69 \pm 0.01 \pm 0.06$ in NSD 
events). For antibaryons we predict \cite{MPRS} increases from $\sim 1.6$ 
for $\bar{p}$ and $\bar{\Lambda}$ to $\sim 2.0$ for $\overline{\Omega}$.

\section{Baryon/antibaryon asymmetry in QGSM}

In the string models baryons are considered as configurations consisting of 
three connected strings (related to three valence quarks) called string 
junction (SJ) \cite{Artru,IOT,RV,Khar}, as it is shown in Fig.~4. Such a 
baryon structure is supported by lattice calculations \cite{latt}. 

\begin{figure}[htb]
\centering
\vskip -2.cm
\includegraphics[width=.5\hsize]{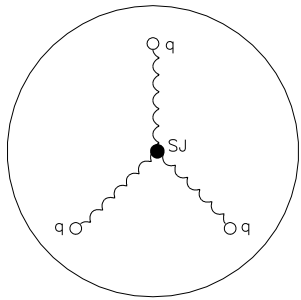}
\vskip -1.2cm
\caption{\footnotesize
The composite structure of a baryon in string models. Quarks are shown by open 
points and SJ by black point.}
\end{figure}

This picture leads to some general phenomenological predictions. In
particular, it opens room for exotic states, such as the multiquark bound 
states, 4-quark mesons, and pentaquarks \cite{RV,DPP1,RSh}. In the case of 
inclusive reactions, the baryon number transfer to large rapidity distances in 
hadron-nucleon reactions can be explained \cite{ACKS} by SJ diffusion.

The production of a baryon-antibaryon pair in the central region usually 
occurs via $SJ$-$\overline{SJ}$ pair production. Then, the SJ, that has upper
color indices, combines with sea quarks, while the antiSJ ($\overline{SJ}$),
that has lower indices, combines with sea antiquarks, resulting, respectively,
into the $B$ and the $\bar{B}$ of a $B\bar{B}$ pair~\cite{RV,VGW},
as it is shown in Figure~5a. 

\begin{figure}[htb]
\centering
\includegraphics[width=.6\hsize]{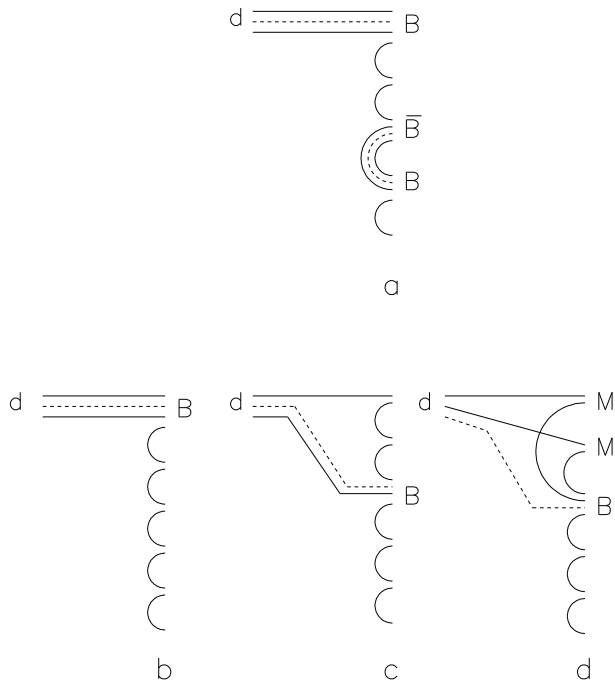}
\caption{\footnotesize
QGSM diagrams describing secondary baryon $B$ production by diquark $d$.
(a) Central production of $\overline{B}B$ pair via production of new
SJ-$\overline{SJ}$ pair. Single production of $B$ in the processes  
of diquark fragmentation: (b) initial SJ together with two valence quarks and 
one sea quark, (c) initial SJ together with one valence quark and two sea 
quarks, and (d) initial SJ together with three sea quarks. Quarks 
are shown by solid curves and SJ by dashed curves.}
\end{figure}

In processes with incident baryons, e.g. in $pp$ collisions, another
possibility to produce a secondary baryon in the central region exists.
This second possibility is connected with the diffusion in 
rapidity space of a SJ existing in the initial state and it leads to 
significant differences in the yields of baryons and antibaryons in the 
midrapidity region even at rather high energies~\cite{ACKS}. Probably, 
the most important experimental fact in favour for this process is the 
rather large asymmetry in $\Omega$ and $\overline{\Omega}$ baryon
production in high energy $\pi^-p$ interactions~\cite{ait}. 

The quantitative theoretical description of the baryon number transfer 
via SJ mechanism was suggested in the 90's and used to predict~\cite{KP1} the 
$p/\bar{p}$ asymmetry at HERA energies. The quantitative description of the
baryon number transfer due to SJ diffusion in rapidity space was first
obtained in \cite{ACKS} and then in papers \cite{AMPS,MRS,MPRS,BS,AMS,Olga}. 

To obtain the net baryon charge we consider three different possibilities
according to ref.~\cite{ACKS}. The first one is the fragmentation of 
the incoming diquark giving rise to a leading baryon (Figure~5b). A second
possibility is to produce a leading meson in the first break-up of the
string and one baryon in a subsequent break-up (Figure~5c). In these two
first cases the baryon number transfer is possible only for short
distances in rapidity. In the third case, shown in Figure~5d, both initial
valence quarks in the diquark recombine with sea antiquarks into mesons
$M$, while a secondary baryon is formed by the SJ together with three sea
quarks. 

The fragmentation functions for the secondary baryon $B$ production 
corresponding to the three processes shown in Figs.~5b-d can be written 
as follows (see~\cite{ACKS} for more details):
\begin{eqnarray}
G^B_{qq}(z) &=& a_N\cdot v^B_{qq} \cdot z^{2.5} \;, \\
G^B_{qs}(z) &=& a_N\cdot v^B_{qs} \cdot z^2\cdot (1-z) \;, \\
G^B_{ss}(z) &=& a_N\cdot\varepsilon\cdot v^B_{ss} \cdot z^{1 -
\alpha_{SJ}}\cdot (1-z)^2  \;,
\end{eqnarray}
for Figs.~5b, 5c, and 5d, respectively, and where $a_N$ is the normalization 
parameter  that was determined from the experimental data at fixed target 
energies, and $v^B_{qq}$, $v^B_{qs}$, $v^B_{ss}$ are the relative probabilities 
for different baryons production that can be found by simple quark 
combinatorics \cite{AnSh,CS}. 

The contribution shown in Figure~5d is essential if the intercept of the SJ 
exchange Regge-trajectory, $\alpha_{SJ}$, is large enough. The contribution 
of the graph in Figure~5d is weighted in QGSM by the coefficient $\varepsilon$ 
which determines the small probability for such a baryon number transfer 
to occur. Let's finally note that this process can be very naturally 
realized in the quark combinatorial approach~\cite{AnSh} through the 
specific probabilities of a valence quark recombination (fusion) with 
sea quarks and antiquarks.

At high energies the SJ contribution to the inclusive cross section 
of secondary baryon production at large rapidity distance $\Delta y$ from 
the incident nucleon can be estimated as
\begin{equation}
\frac{1}{\sigma}\cdot\frac{d\sigma^B}{dy} \sim a_B\cdot\varepsilon\cdot
e^{-(1 - \alpha_{SJ})\cdot\Delta y} \;,
\end{equation}
where $a_B = a_N\cdot v^B_{ss}$. 

The data by the ALICE Collaboration \cite{ALI} for $\bar{p}/p$ ratios in 
$pp$ collisions at LHC energies $\sqrt{s} = 900$ GeV and 7 TeV are presented 
in Table 4, together with QGSM the corresponding calculations with different
values of $\alpha_{SJ}$. These data seem to favour the value
$\alpha_{SJ} = 0.5$, what could mean that the Odderon contribution
is not seen \cite{MRS} in this process.

\begin{center}

\vskip 5pt

\begin{tabular}{|c||r|r|} \hline
SJ exchange & $\sqrt{s} = 900$ GeV & $\sqrt{s} = 7$ TeV \\ \hline
$\alpha_{SJ} = 0.9$ & 0.89\hspace{1.cm} & 0.95\hspace{1.cm} \\
$\alpha_{SJ} = 0.5$ & 0.95\hspace{1.cm} & 0.99\hspace{1.cm} \\
$\varepsilon = 0$\hspace{0.2cm}  & 0.98\hspace{1.cm} & 1.\hspace{1.4cm}   
\\ \hline
ALICE~\cite{ALI} & $0.957 \pm 0.006 \pm 0.014$ & $0.991 \pm 0.005 \pm 0.014$ 
\\ \hline
\end{tabular}

\end{center}
Table 4. {\footnotesize The QGSM predictions for $\bar{p}/p$ in $pp$
collisions at LHC energies and the data by the ALICE
Collaboration \cite{ALI}. The value $\varepsilon = 0$ corresponds
to the case with not C-negative exchange.} 

\vskip 5pt

The LHCb Collaboration measured the ratios of $\bar{\Lambda}$ to $\Lambda$ 
in the rapidity interval $2 < y < 4$ at $\sqrt{s} = 900$ GeV and 7 TeV
\cite{LHCb}. We compare these results with the QGSM calculations in Fig. 6. 

\begin{figure}[htb]
\centering
\includegraphics[width=.49\hsize]{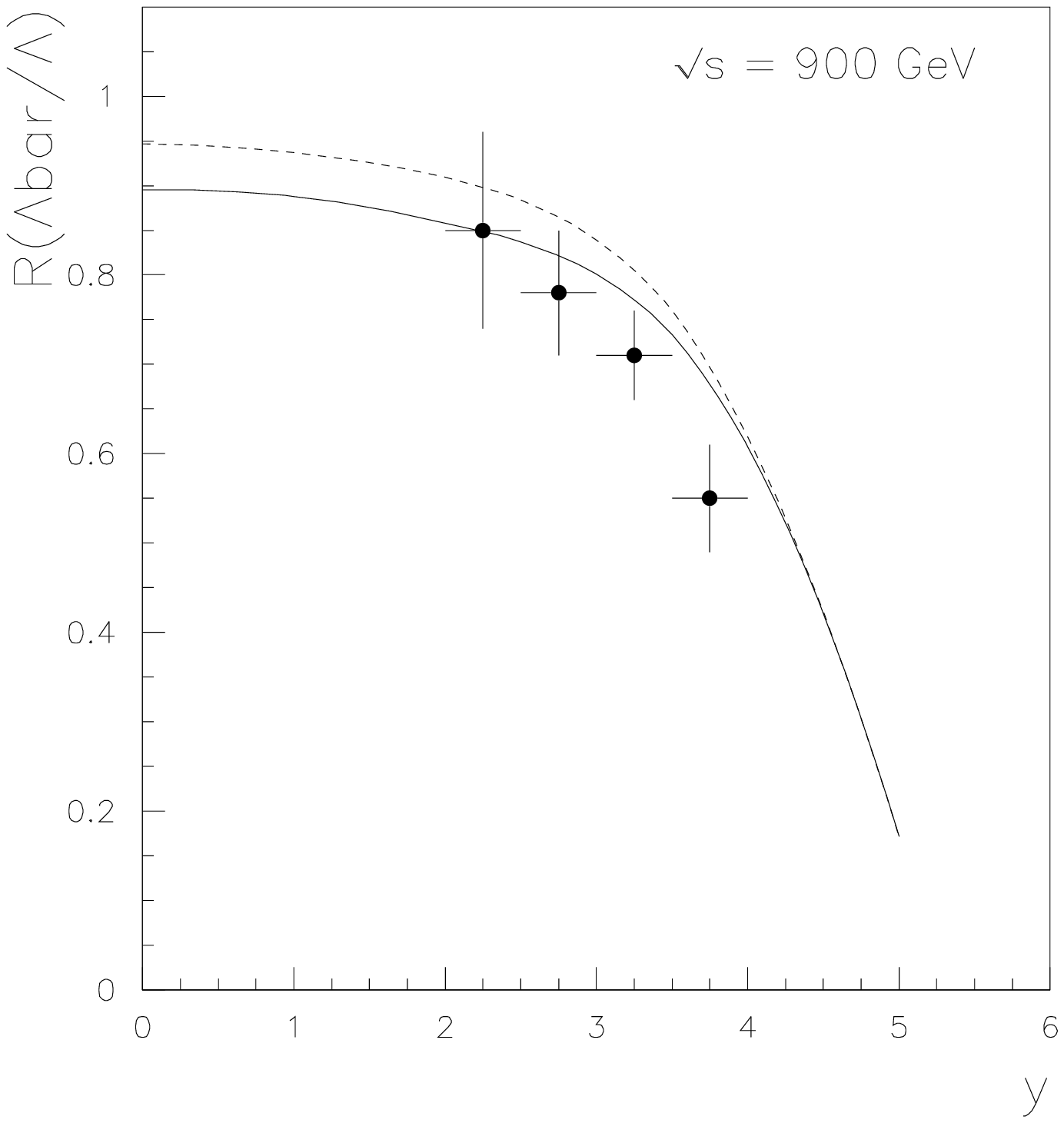}
\includegraphics[width=.49\hsize]{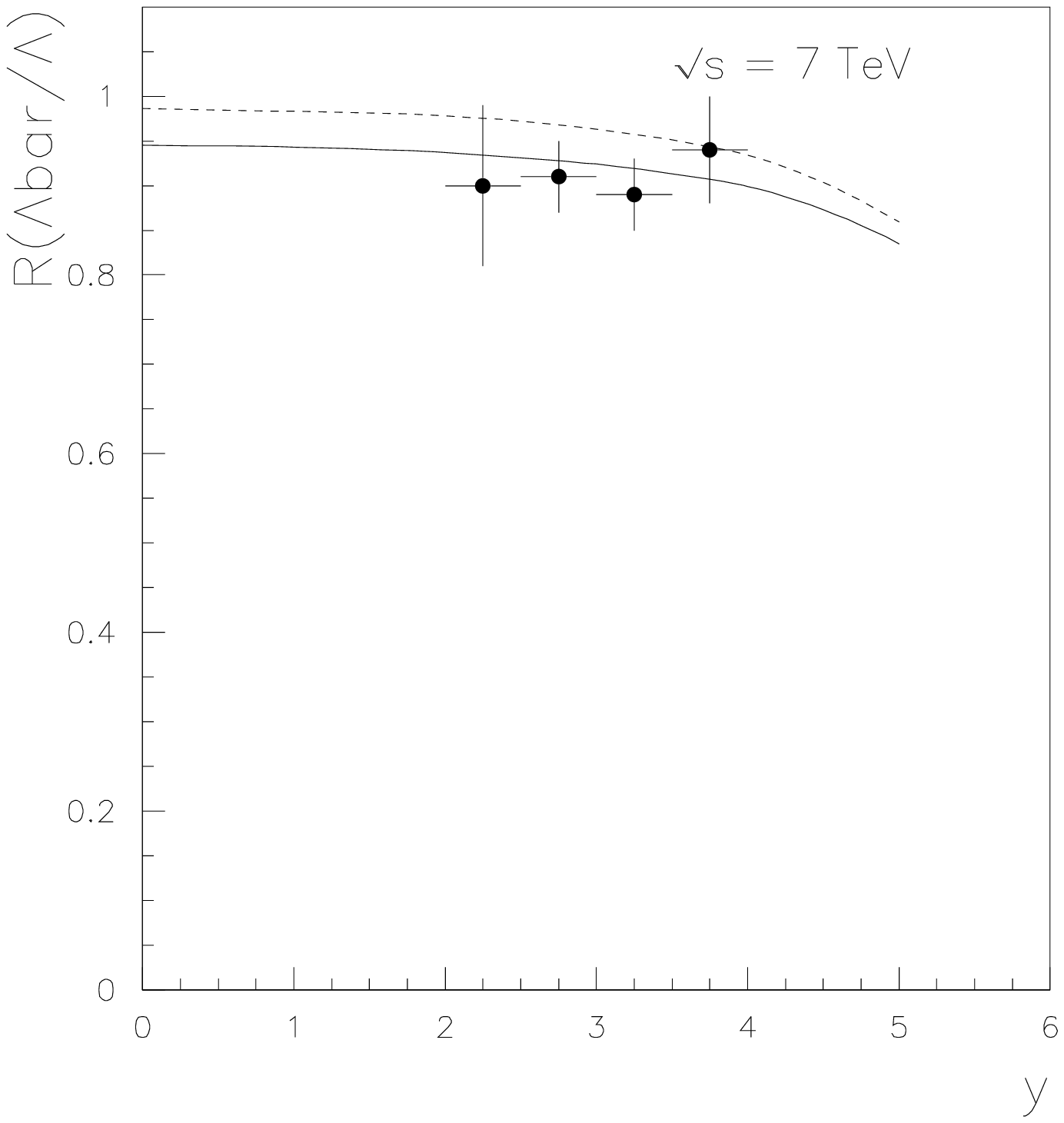}
\caption{\footnotesize 
The QGSM predictions (solid curves for $\alpha_{SJ} = 0.9$ and dashed
curves for $\alpha_{SJ} = 0.5$) for the ratios of the spectra of secondary 
$\bar{\Lambda}$ to $\Lambda$ as the functions of their rapitities 
at energies $\sqrt{s} = 900$ GeV (left) and $\sqrt{s} = 7$ TeV  
(right), together with the experimental data by the LHCb Collaboration 
\cite{LHCb}.}
\end{figure}

One can see that here the QGSM calculation with the value $\alpha_{SJ} = 0.9$
is in a slightly better agreement with the data. It is also necessary to note
that the predictions of PYTHIA 6 MC generator are in disagreement with the
experimental data at $\sqrt{s} = 900$ GeV. 

We predict practically equal $\overline{B}/B$ ratios for baryons with 
different strangeness content.

\section{Conclusion}

The first experimental data obtained at LHC are in general agreement with
the calculations provided by QGSM in the framework of Regge theory with the
same values of parameters that were determined at lower 
energies (mainly for the description of the data of fixed target experiments). 

We neglect the possibility of interactions between Pomerons (so-called 
enhancement diagrams) in the calculations of integrated cross sections and
inclusive densities. Such interactions are very important in the cases of 
heavy ion \cite{CKT} and nucleon-nucleus \cite{MPS} interactions at RHIC 
energies, and their contribution should increase with energy. However, we 
estimate \cite{MPS} that the contributions of these enhanced diagrams to 
the inclusive density of secondaries produced in $pp$ collisions at LHC 
energies is not large enough to be significant.

{\bf Acknowledgements}

We are grateful to $\frame{A.B. Kaidalov}$ for useful discussions and 
comments. This paper was supported by Ministerio de Educaci\'on y Ciencia 
of Spain under the Spanish Consolider-Ingenio 2010 Programme CPAN 
(CSD2007-00042), by Xunta de Galicia project FPA 2005--01963, and, in part, 
by grant RSGSS-3628.2008.2.


\end{document}